\documentclass[11pt,dvips]{article}
\usepackage{amscd,verbatim}
\usepackage[all]{xy}
\usepackage{xypic}
\input xypic
\usepackage{graphicx}
\usepackage{amsmath}
\usepackage[psamsfonts]{amssymb}
\usepackage{amsthm}
\usepackage{euscript}

\usepackage{latexsym}

\renewcommand{\(}{\begin{equation}}
\renewcommand{\)}{end{equation} \vspace{-.05in}\linebreak}

\newcounter{saveeqn}
\newcounter{savealpheqn}

\newcommand{\alpheqn}{\setcounter{saveeqn}{\value{equation}}%
  \stepcounter{saveeqn}\setcounter{equation}{0}%
  \renewcommand{\theequation}{\mbox{\arabic{section}.\arabic{saveeqn}
\alph{equation}}}
  \renewcommand{\)}{\end{equation}}}
\def\part#1{\frac{\partial}{\partial{#1}}}%
\def\group#1{\refstepcounter{equation}\setcounter{saveeqn}{\value{equati 
on}}%
  \label{#1}\setcounter{equation}{0}%
\renewcommand{\theequation}{\mbox{\arabic{section}.\arabic{saveeqn}
\alph{equation}}}
  \renewcommand{\)}{\end{equation}}}
\newcommand{\reseteqn}{\setcounter{equation}{\value{saveeqn}}%
  \renewcommand{\theequation}{\arabic{section}.\arabic{equation}}%
  \renewcommand{\)}{\end{equation}}}

\newcommand{\aalpheqn}{\setcounter{saveeqn}{\value{equation}}%
  \stepcounter{saveeqn}\setcounter{equation}{0}%
  \renewcommand{\theequation}{\mbox{
        \Alph{subsection}.\arabic{saveeqn}\alph{equation}}}
   \renewcommand{\)}{\end{equation}}}
\newcommand{\areseteqn}{\setcounter{equation}{\value{saveeqn}}%
  \renewcommand{\theequation}{\Alph{subsection}.\arabic{equation}}%
  \renewcommand{\)}{\end{equation}}}

\renewcommand{\thefootnote}{\alph{footnote}}
\renewcommand{\(}{\begin{equation}}
\renewcommand{\)}{\end{equation}}
\newcommand{\ba}{\begin{eqnarray}}
\newcommand{\ea}{\end{eqnarray}}

\renewcommand{\r}{\rho}
\newcommand{\bp}{\mathop{\vtop{\ialign{##\crcr
   $\hfil\displaystyle{}\hfil$\crcr\noalign{\kern-13pt\nointerlineskip}
   \BIG{(}\hskip0pt\crcr\noalign{\kern3pt}}}}}
\newcommand{\cbp}{\mathop{\vtop{\ialign{##\crcr
   $\hfil\displaystyle{}\hfil$\crcr\noalign{\kern-13pt\nointerlineskip}
   \BIG{)}\hskip0pt\crcr\noalign{\kern3pt}}}}}
\newcommand{\pa}{\mathop{\vtop{\ialign{##\crcr
    
$\hfil\displaystyle{\oplus}\hfil$\crcr\noalign{\kern+1pt\nointerlineskip 
}
   \hspace{.08in}$^{\alpha=0}$\hskip6pt\crcr\noalign{\kern3pt}}}}}

  {\renewcommand{\theequation}{\Alph{subsection}.\arabic{equation}}%
   \renewcommand{\thesubsection}%
                {Appendix \Alph{subsection}.\setcounter{equation}{0}}%
   \renewcommand{\alpheqn}{\aalpheqn}%
   \renewcommand{\reseteqn}{\areseteqn}
   }

\newcommand{\R}{\ensuremath{\mathbb R}}

\newcommand{\C}{\ensuremath{\mathbb C}}

\newcommand{\Z}{\ensuremath{\mathbb Z}}

\def\r{\rightarrow}

\newcommand{\beq}{\begin{equation}}
\newcommand{\beg}[2]{\begin{equation}\label{#1}#2\end{equation}}
\newcommand{\eeq}{\end{equation}}

\numberwithin{equation}{section}

\def\hsp#1{\hspace{#1in}}

\newcommand{\rref}[1]{(\ref{#1})}

\catcode`\@=11
\def\vereq#1#2{\lower3pt\vbox{\baselineskip1.5pt \lineskip1.5pt
\ialign{$\m@th#1\hfill##\hfil$\crcr#2\crcr\sim\crcr}}}
\catcode`\@=12

\makeatletter
\newcommand\figcaption{\def\@captype{figure}\caption}
\newcommand\tabcaption{\def\@captype{table}\caption}
\makeatother
\renewcommand{\(}{\begin{equation}}
\renewcommand{\)}{\end{equation}}

\newcommand{\CC}{{\mathbb C}}
\newcommand{\RR}{{\mathbb R}}
\newcommand{\ZZ}{{\mathbb Z}}
\newcommand{\QQ}{{\mathbb Q}}

\theoremstyle{plain}

\theoremstyle{definition}

\begin{document}

\begin{titlepage}
\begin{flushright}


hep-th/0410293
\end{flushright}

\vspace{2em}
\def\thefootnote{\fnsymbol{footnote}}

\begin{center}
{\Large\bf Type IIB string theory, S-duality, and generalized 
cohomology}
\footnote{I. K. is supported by NSF  
grant DMS 0305853, and H. S. is supported by the Australian Research 
Council.} 
\end{center}
\vspace{1em}
\begin{center}
   Igor Kriz\footnote{E-mail: \tt ikriz@umich.edu}$^1$ and
Hisham Sati \footnote{E-mail: \tt
hsati@maths.adelaide.edu.au}$^{2,3}$

\end{center}

\begin{center}
\vspace{1em}
{\em  { $^1$Department of Mathematics\\
            University of Michigan\\
            Ann Arbor, MI 48109,\\ 
            USA\\
\hsp{.3}\\
$^2$Department of Physics\\
       University of Adelaide\\
       Adelaide, SA 5005,\\ 
       Australia}\\
\hsp{.3}\\
$^3$Department of Pure Mathematics\\
       University of Adelaide\\
       Adelaide, SA 5005,\\ 
       Australia}\\

\end{center}  

\vspace{0em}

\begin{abstract}
In the presence of background Neveu-Schwarz flux,
the description of the Ramond-Ramond fields of type IIB string theory 
using twisted $K$-theory is not compatible with S-duality.
We argue that other possible variants of twisted K-theory would still not
resolve this issue. We propose instead a 
connection of S-duality with elliptic cohomology,
and a possible T-duality relation of this to a previous proposal
for IIA theory, and higher-dimensional limits.
In the process, we obtain some other results which may
be interesting on their own. In particular, we prove a conjecture of
Witten that the $11$-dimensional spin cobordism group vanishes on $K(\Z,6)$,
which eliminates a potential new $\theta$-angle in type IIB string theory.

\end{abstract}

\vfill

\end{titlepage}
\setcounter{footnote}{0}
\renewcommand{\thefootnote}{\arabic{footnote}}
\newcommand{\cform}[3]{\begin{array}{c}
{\scriptstyle #3}\\
#1\\
{\scriptstyle #2}\end{array}}

\pagebreak

\renewcommand{\thepage}{\arabic{page}}

\section{Introduction}
There are two string theories with two supersymmetries in ten dimensions,
nonchiral type IIA and chiral type IIB theories. Type IIB supergravity
\cite{GS} is the classical low energy limit of type IIB string theory. 
There is no manifestly Lorentz-invariant action for this theory 
\cite{Mar}, but one can write down the equations of motion 
\cite{Schw},\cite{HW}, and the symmetries and transformation rules 
\cite{Sch-W}. An action which encodes self-duality was proposed in
\cite{PST}. However, such an action does not seem to be well-adapted
for the topologically nontrivial situations we are interested in.
Type IIB theory is free of chiral anomalies \cite{Alv-W}.

\vspace{3mm}
Since both are ten-dimensional, why is IIB different than IIA? First
note that in both cases, what is self-dual is the total field strength
$F=\sum_i F_i$ with $i=2p$ for IIA and $i=2p+1$ for IIB.
In the IIA story, $F_4$ might be viewed as having somewhat of a 
more transparent role than the other fields because that is
the field directly descending from the fundamental field of M-theory, 
namely
$G_4$. In IIB, there is no obvious way of picking one of the field
(because it is not as transparently related to M-theory), but it
should have a rank close to four, i.e. either the three-form
or the five-form. Dualities give some hints. M-theory on a torus $T^2$
is equivalent to type IIB on a circle $S^1$ in the limit when the
volume of the torus shrinks to zero (i.e. T-duality) \cite{Asp} 
\cite{SchM}. So if we start     
with $G_4$ in M-theory, then we can either reduce diagonally
or vertically. The result of the former is a two-form, $F_2$ which can 
only
be lifted over the circle diagonally to a three-form of type IIB. For the 
latter, we still have a four-form $F_4$ which can only lift diagonally
to the IIB five-form. So in this way we see that the fields of
degrees three and five in IIB are what is taking role which $F_4$ played 
in IIA. A detailed account of the relation between IIB and M-theory
can be found in \cite{Gr}.

\vspace{3mm}
Self-duality is usually a subtle issue to deal with when quantizing 
(see e.g. \cite{Hen} \cite{Wi4}),
via the path integral in our case. In type IIA, this was taken care
of in \cite{DMW} by summing over a set of commuting periods for the 
partition
function, which amounted to picking a polarization over the torus
$K^0/K_{tors}^0$ which keeps only the forms $F_0$, $F_2$ and $F_4$ and 
kills 
the ``duals''. Then one could still further ignore all but $F_4$. Now 
the problem with just $F_4$ makes sense quantum-mechanically, because
for one thing, it is not self-dual, and so the partition function is
well-defined.
In type IIB, a'priori $F$ is not a sum of
even number of terms as was the case in IIA, where had the $\ZZ_2$-graded
expansion
\(
F=F_0 + F_2 + F_4 + F_6 + F_8 + F_{10}.
\)
Now in IIB we have
\(
F=F_1 + F_3 + F_5 + F_7 + F_9
\) 
which contains only five terms. So it is not obvious how to pick
a polarization which kills half the terms as was done in IIA.
However, one might argue that $F_5$ can be split into two parts according 
to
self-duality/antiselfduality. One might also argue that since the theory 
has a 
(spacetime-filling) 
$D9$-brane, which couples to  
a 10-form potential $C_{10}$, one requires a corresponding field 
strength
$F_{11}$. But in 10-dimensions this cannot occur. However, the 
rescue might be in looking at higher dimensions. If we take the view that 
type IIB theory has an intrinsic 12-dimensional
structure associated to it, by viewing the $SL(2,\ZZ)$ S-duality
symmetry of the moduli, as actually coming from a physical torus on
which the theory had been compactified, then one can identify the
two modular groups and one gets a 12-dimensional lift of IIB, namely
F-theory \cite{Vafa}. This is motivated by the fact that the S-duality 
group
$SL(2,\ZZ)$ is in fact a U-duality group \cite{HT} and the use of the  
intuition 
from
compactification of M-theory to ``discover'' this as the modular group
of an internal torus. If we take this point of view, then we have
a total RR field strength with an even number of terms and thus we
can pick a polarization that kills $F_{2p+1}$ for $p\geq 3$.

\vspace{3mm}
Next, one has to pick whether to keep $F_3$ or $F_5$, and further
worry about analyzing the problem in an
S-duality-covariant fashion. This is implied in the
self-duality of $F_5$, which as we mentioned above, will have to
be treated very carefully in the partition funcion,
and
the S-duality symmetry of $F_3$ which amounts to having to
consider $H_3$ at an equal footing, that is look at the pair as an
$SL(2,\ZZ)$-doublet, i.e. as transforming in the two-dimensional
representation of (subgroup of) the modular group.

\vspace{3mm}
Indeed, one of the problems that was discussed in \cite{DMW} is that 
there seems
to be no symmetry between the Neveu-Schwarz three-form $H_3$ and the 
Ramond-Ramond three-form $F_3$, because traditionally $F_3$ is viewed as
a $K$-theory object whereas $H_3$ is an ordinary differential form.
They propose two approaches: The first is that there is a source term
$Q$ for the anomaly equation, so that
\footnote{At the level of differential forms, i.e. over $\RR$ or $\CC$,
this term is a wedge product of two copies of the same odd differential
form, and thus vanishes identically. However, over $\ZZ$, this can be 
$2$-torsion. Note that
S-duality and U-duality care about integral information.} 
 $Q=H_3\cup H_3$.
The second is that the $SL(2,\ZZ)$ invariance of the theory does not
come from a simple transformation law on the space of classical
configurations, i.e. the invariance of the partition function need not
come from an invariance on the classical fields but appears only
when one looks at equivalence classes (as was done in comparing M-theory
to IIA). However, no convincing scenario for the latter was found.
\vspace{3mm}

The above can be rephrased in terms of differentials in the 
Atiyah-Hirzebruch Spectral Sequence (AHSS). In particular, the 
problem is to find an S-duality covariant extension of the 
third differential. Some discussions of this issue has appeared
in \cite{EV}. For the twisted case, without the S-duality issue,
the third differential was studied and given a very
nice physical interpretation in \cite{MMS}. From the point of view 
of branes and solitons, some further discussion on this can be found in
\cite{J2} \cite{J1}. Some progress in the
resolution of the S-duality problem was made in \cite{DFM} from
a different point of view, using their model of the M-theory 
three-form.

\vspace{3mm}
Now let us take this one step further. If one takes the view that
$H_3$ is interpreted as a twisting for $K$-theory (\cite{Wi1} and 
\cite{Kap} \cite{BM} \cite{HM}), then $H_3$ is more naturally 
associated to a gerbe than a differential form. A gerbe is in some 
sense a ``higher'' object than that of $K$-theory, i.e. a vector bundle,
and so one can argue that in order to provide a description that 
treats the NS fields and the RR fields on equal footing, one has
to go beyond $K$-theory.  

\vspace{3mm} 
The question then is which (twisted) generalized cohomology theory 
is the right one to use? In type IIA, the answer \cite{KS} was 
somewhat more manifest
since the Diaconescu-Moore-Witten anomaly \cite{DMW} $W_7$ was given 
by $\beta Sq^2(x_4)$, where $x_4$ is a four-class which could naturally be 
identified with four-form. In type IIB, however, there is no natural
four-class available and so the situation is a little bit subtle. 

\vspace{3mm}
In this paper, we find that to resolve the S-duality problem in IIB, 
one cannot use any variant of twisted $K$-theory where the twisting
is merely by an integral cohomology class in dimension $3$. Such theory
would be classified by a fibered space
over the space classifying $K^1$ whose fiber would be $K(\Z,3)$, but 
we find that there is no such space in which the first Postnikov invariant
would be compatible with S-duality. It is important to point out that
our analysis does {\em not} exclude the possibility of a solution which
combines the $3$-dimensional twisting with some higher twisting. 
On the other hand, we note that such higher twisting would amount to
constructing, term by term, a Postnikov tower of some space which merely maps
to $K^1$. We include some speculation about the nature of that space
or generalized cohomology theory.

\vspace{3mm}
More precisely, the authors have a candidate theory which they
found in connection with the T-dual scenario of IIA in \cite{KS},
namely {\em elliptic cohomology}. What happens is that under certain
conditions, there exists an elliptic cohomology
analogue of the physical theory which defines the
type II partition functions via $K$-theory \cite{DMW, Wi4}.
Elliptic cohomology, by design, has built in modularity which
seems indeed to correspond with IIB S-duality. 
It is possible that in a suitable scenario, the sources $F_3$ and
$H_3$ would both originate in elliptic cohomology which might resolve
the puzzle in that range. We do not claim, however, that such scenario occurs 
in classical weak-coupled IIB string theory. Although IIB theory is
U-dual to itself, it is nevertheless possible that on a line in moduli
space connecting the two U-duals, dimensional expansion may occur.
Dimensional expansion should, in fact, be generic. We think that it is
that range to which the elliptic partition function corresponds, and
we will pursue this direction in follow-up work.

\vspace{3mm}
There is another piece of evidence that the elliptic cohomology partition
function should involve dimensional expansion and unification of string
theories.
The existence of the elliptic theory in the untwisted case requires the condition
$w_4(X^{10})=0$, which is a stronger condition than the untwisted IIA obstruction
$W_7(X^{10})=0$ (we will see that in untwisted IIB theory, there is no obstruction
at all; that doesn't violate T-duality, as the $W_7$-obstruction vanishes on
a manifold of the form $S^1\times X^{9}$). The appearance of the $4$-dimensional
obstruction may be somewhat surprizing. However, such obstruction occurs in
the other types of string theory, namely type I and heterotic, where it relates
the $w_4$ to the $4$-form of the gauge field. If the gauge field is $0$, we get
precisely the obstruction $w_4=0$. We think that in sector in which our elliptic
partition function applies, type II string theory becomes unified with other
string theories, which is the reason why we are seeing a unified obstruction.
More precise proposals will be made in follow-up work.

\vspace{3mm}
This paper is organized as follows. In section \ref{CS} we study the 
Chern-Simons 
construction for the five-form, and we prove the vanishing fo the 
spin cobordism group $\tilde{\Omega}_{11}^{spin}\left( K(\ZZ,6)\right)$,
which was conjectured by Witten. 
This fact, in fact, is interesting on its own, as it eliminates another
potential $\theta$-angle in type IIB string theory.
We also
comment on the construction of the intermediate Jacobian for cohomology,
making a connection to the work of Hopins and Singer.

\vspace{3mm}
In section \ref{K1} we review Witten's construction of the RR partition 
function
in cohomology and in $K$-theory, and provide some details for the case of
$H_3$-twisted $K$-theory.

\vspace{3mm}
In section \ref{DMWa} we give the argument that there is no 
Diaconescu-Moore-Witten anomaly in ordinary
untwisted $10$-dimensional type IIB partition function. 

\vspace{3mm} 
In section \ref{S} we discuss the main topic which is the S-duality puzzle 
and previous attempts and suggestions to resolve it, and also provide
homotopy-theoretical evidence why a straightforward solution to
the S-duality puzzle, using some generalized twisting of $K$-theory,
does not seem possible.

\vspace{3mm}
In section \ref{vs}, we propose in detail our possible
approach to the S-duality
problem using elliptic cohomology, and also write down the $\theta$-bundle
construction, which is the main step of the partition function construction,
in this elliptic cohomology context. This 
is a unification of the IIA elliptic cohomology
partition function correction of \cite{KS} and the IIB $K^1$-partition 
function construction of \cite{Wi4}.

\vspace{3mm}
Having proposed a solution to the S-duality problem, one must address 
T-duality.
In \cite{KS} we found that the cancellation of the Diaconescu-Moore-Witten
anomaly in type IIA required the formulation of the partition function
in terms of elliptic cohomology. How are the two pictures compatible?
We address this question in section \ref{T}.

\section{The partition function and the Chern-Simons construction}
\label{CS}

We start by reviewing Witten's construction.
In order to study the partition function, one has to integrate
over the space of fields. At the level of cohomology, one would also
like to interpret the partition function as a section of a line bundle
${\mathcal L}$ over the intermediate Jacobian $J^5$, such that the first
Chern class $c_1({\mathcal L})$ equals the polarization, or symplectic
form, $\omega$ of $J^5$. Witten showed that using the Chern-Simons 
construction
one can always find a line bundle with $c_1=2\omega$, and further, that
in order to get one with $c_1=\omega$ (i.e. basically using level half)
one has to have, for a closed
twelve-dimensional spin manifold $Z^{12}$, that the
intersection form on $H^6(Z^{12}, \ZZ)$ be even, which is always the case.
\footnote{Recall that in M-theory, the even-ness of the intersection form  
on
the fourth integral cohomology was equivalent to integrality of the
Spin class $\lambda/2$, because of the factor
\(
(-1)^{\int_{\Sigma_4} \lambda} \exp\left(i\int_{\Sigma_4} G_4
\right)
\)
that had to be well-behaved.}

\vspace{3mm}
In the Chern-Simons construction, i.e. extending by a circle with a
bounding spin structure and then
taking a coboundary, one encounters an obstruction for
extending the gauge ``bundles''
as the
cobordism group
$\Omega_{11}^{spin}\left(K(\ZZ,6) \right)$. Witten  
conjectures \cite{eff} 
that this group should be zero because it is unlikely that a theta
angle, parametrized by the dual group
$Hom\left(\Omega_{11}^{spin}\left(K(\ZZ,6) \right), U(1) \right)$,
exists in type IIB string theory.

\vspace{3mm}
Spin cobordism is represented by a spectrum, which is usually 
denoted by $MSpin$. Thus,
another way of writing the above cobordism group, interpreting
it as a generalized homology group using 
the notation of spectra, is $MSpin_{11}K(\ZZ,6)=\Omega_{11}^{spin}(K(\ZZ,6))
$. Recall that by
a result of Milnor, $MSpin_{11}(*)=0$,
so it suffices to consider the reduced group 
$\widetilde{MSpin}_{11}K(\Z,6)=\tilde{\Omega}_{11
}^{spin}(K(\ZZ,6))$. We want to show that
this is zero. The point is that this is in the so-called {\it stable
range}. 

\vspace{3mm}
The word `stable' is used in many different meanings in different contexts.
The algebraic topology meaning, which we are referring to here,
is actually related to the notion of stability of $D$-branes, but we are 
not interested in that at the moment. In algebraic geometry, on the
other hand, for example a `stable bundle' (or curve) is something quite different,
controlling automorphisms. 
In algebraic topology, `stable' means `preserved under suspension'. 
Suspension $\Sigma X$ of a space $X$ is formed by taking the product $X\times [0,1]$
and identifying all the points $(x,0)$ to a single point, and all the points
$(x,1)$ to another single point. Now the crucial property is that 
for any generalized homology theory $E$, we have a natural isomorphism
$$\tilde{E}_n(X)\cong \tilde{E}_{n+1}\Sigma X$$
where $\tilde{E}_n$ denotes the generalized homology group. A similar relation
holds also for cohomology. This is what we mean by saying that generalized
homology (or cohomology) theories are stable. In physics, it is important
that stable D-branes are classified by $K$-theory, i.e. a stable invariant.
The two notions of stability are related, but as we already commented, we
are not using that in this proof.

\vspace{3mm}
Now conversely, with very mild restrictions, essentially every stable invariant
of spaces is represented by a generalized homology or cohomology theory.
This means that many common invariants are not stable. Homotopy groups are
one example. Nevertheless, there is always a natural homomorphism
(called the suspension map)
\beg{efreud}{\pi_n(X)\r \pi_{n+1}\Sigma X}
and the Freudenthal suspension theorem asserts that \rref{efreud} is in
fact an isomorphism if $n\leq 2k$ and onto for $n=2k+1$ where $k$ is such number that 
$$\pi_i(X)=0$$
for $i\leq k$ (we say that the space $X$ is {\em $k$-connected}). Because
\rref{efreud} is an isomorphism in this range of $n$, we refer to $n\leq 2k$ as
the {\em stable range}. Note that by mere comparison of numbers, once $n$
is in the stable range for $X$, then every suspension map
$$\pi_{n+i}\Sigma^i X\r \pi_{n+i+1}\Sigma^{i+1}X$$
is also an isomorphism for $i\geq 0$. 

\vspace{3mm}
This led algebraic topologists to realize that there is a functor $\Sigma^{\infty}$
of `infinite suspension' from spaces to the `stable world' i.e. generalized
cohomology theories. For these purposes, actually, it is useful to think
of generalized cohomology theories on a point set level, not just up to homotopy.
From this point of view, they are referred to as spectra. The essential point
is that for any generalized homology $E$, we have a natural isomorphism
$$\tilde{E}_n X\cong E_n \Sigma^{\infty}X$$
because generalized homology is a stable invariant! A similar relation
also holds for generalized cohomology.

\vspace{3mm}
Now we have a map of spectra
\(
\Sigma^{\infty} K(\ZZ,6) \rightarrow \Sigma^6 H\ZZ.
\label{spec}
\)
Here the spectrum on the right hand side is integral homology suspended
by $6$. This means that when applying the homology theory represented by this
spectrum to a space, the $i$'th integral homology group appears in dimension
$i+6$. The map \rref{spec} comes simply from the fact that the $6$'th
integral cohomology group of $K(\Z,6)$ is $\Z$. 
But by the Freudenthal suspension theorem, we can say what the map \rref{spec}
does in homotopy groups. This is because $K(\Z,6)$ is $5$-connected
(and in fact its only non-trivial homotopy group is $\Z$ in
dimension $6$), so
by the Freudenthal suspension theorem we know that $\Sigma^{\infty} K(\Z,6)$ has
no non-trivial homotopy groups, except $\Z$ in dimension $6$, in dimensions
$\leq 11$. 
It follows that the map \rref{spec} induces on homotopy groups an isomorphism
in dimension $\leq 11$, and an onto map in dimension $12$ (simply because
the right hand side has no homotopy in that dimension). Such map 
is referred to as a $12$-equivalence. In general, an $n$-equivalence
of spaces or spectra has the property that it induces an isomorphism of any
connective generalized homology or cohomology theory in dimension $<n$, and an onto
map in dimension $n$ (connective means that the negative generalized homology
resp. cohomology groups of a point are $0$). 
This is a consequence of the Whitehead theorem.

\vspace{3mm}
Therefore, \rref{spec}, which is a 12-equivalence, induces iso in
dimensions less than twelve and onto map in dimension twelve in any
connective generalized homology theory, of which $MSpin$ is an example.
So all we need is
\(
MSpin_{11}\Sigma^6 H\ZZ =H_5(MSpin, \ZZ)
\label{epspin}
\)
(This is an interesting property of spectra: they are simultaneously
generalized homology theories, and their arguments. Further, 
for spectra $E,F$, we have $E_n F=F_n E$. We refer the reader to \cite{adams}
for this fact, as well as basic foundations of the theory of spectra,
which should include all the properties used in this proof.)
But now by Thom isomorphism, the right hand side
of \rref{epspin} is equal to $H_5(BSpin, \ZZ)$, which is
zero.
\footnote{Note that the calculation of the other groups that occur
in Type IIA and M-theory, such as $MSpin_{i} K(\ZZ,4)$ for $i=10,11$
can be understood with the same philosophy. However, there one has to be
careful in seperating {\it stable} and {\it unstable} parts, which
correspond to additive and non-additive structures.}

\vspace{3mm}
We will next comment on the construction of the Jacobian over the
integral cohomology torus. Recall that the theory has, in the RR sector,
the field strengths $F_{2p+1}$, $p\leq 0 \leq 4$. In the case of trivial
cohomology, $F_{2p+1}=dC_{2p}$.
For $p=2$ the field strength is self-dual $F_5=*F_5$. This is
analogous to the story of a self-dual scalar in 2 dimensions and the
chiral 2-form theory on the M5 worldvolume \cite{eff} \cite{Wi4}. These
are similar because
of chirality in dimensions $4k+2$. The construction of the
intermediate Jacobian (of cohomology)
of the string and the fivebrane worldvolume theories should then be
helpful for the type IIB case. Indeed this is the case as the work
of Hopkins and Singer \cite{HS} suggests. Their construction of the
line bundle works for any $Spin$-manifold M of dimension $4n-2$.
Note that in order for their monodromy formula to work, they
require that the cobordism group $MSpin_{4n-1}K(\ZZ,2n)$
( $= H_{2n-1}(BSpin,\ZZ)\oplus MSpin_{4n-1}(*)$ as above) vanishes. Note that in our
case, $n=3$, we have proved above the vanishing of the corresponding
cobordism group, and thus at the level of cohomology one can get
the corresponding line bundle by applying the formalism of \cite{HS}.

\vspace{3mm}
Note that in the special case when $X=T^{10}$, Dolan and Nappi \cite{DN}
found the $SL(10,\ZZ)$-invariant partition function for the
chiral five-form. However, as they also indicate, their methods would not 
extend to manifolds with more nontrivial topology.

\section{$K^1$ in type IIB}
\label{K1}
First we start by reviewing the appearance of $K$-theory in type IIB.
The $RR$ fields of type IIB, in the absence of branes and $NSNS$
fields, are determined by $K^1(X^{10})$ \cite{Wi1}. In the presence of
$NSNS$ field $H_3$ then this was suggested in \cite{Wi1}, and shown in
\cite{Kap} for the torsion case and later in \cite{BM} for nontorsion,   
to be the
twisted group $K_H^1(X^{10})$. The formula for the $RR$ field strengths
was proposed in \cite{MW} to be the same as for type IIA theory,  
\(
\frac{F(x)}{2\pi}=\sqrt{{\hat A}(X)}ch(x)
\)
where now the element $x$ has a slightly different interpretation.
Since $K^1(X)$ is isomorphic to ${\widetilde K}^0(X \times S^1)$
\footnote{the reduced group, i.e. the subset of $K^0(X \times S^1)$
consisting of elements that are trivial when restricted to $X$.}
, the
Chern
character $ch(x)$ is an element of $H^{even}(X \times S^1)$, which, upon
integration over $S^1$, maps to $H^{odd}(X)$. One aspect of the
$K$-theory interpretation of the $RR$ fields is that the torsion parts
are also unified and one cannot make sense of torsion parts of seperate
components of $F(x)$.

\vspace{3mm}
The construction of the theta functions was explained in \cite{MW} relying
on
the definitions in \cite{Wi4}. At the level of cohomology, on a
ten-dimensional spin manifold, the potential
$C_4$ with a
self-dual field strength $F_5$, is topologically classified by 
 a class $x \in H^5(X, \ZZ)$, which is represented in de Rham cohomology
by $F_5/{2\pi}$. The partition function is then constructed by summing  
over
all values of $x$.
This happens when one ignores the self-duality and
imposes the conventional Dirac quantization.
However, for a self-dual $F_5$, the partition function is not
obtained by summing independently over all values of $x$, but rather
\cite{Wi4} in terms of a theta function on ${\mathcal T}=H^5(X,U(1))$,
which, in the
absence of torsion in the cohomology of $X$, is the torus
$H^5(X, \RR)/H^5(X,\ZZ)$. Then the theta function is constructed by     
summing over a maximal set of ``commuting'' periods.
In order to determine the line bundle over ${\mathcal T}$, whose
section is the partition function, we need to detemine the
the $\ZZ_2$-valued function $\Omega(x)$ on $H^5(X,\ZZ_2)$. As in the
IIA case, this function obeys (for all $x$, $y \in H^5(X,\ZZ_2)$)
\(
\Omega(x+y)=\Omega(x) \Omega(y) (-1)^{x\cdot y}
\)
where $x\cdot y$ is the intersection pairing on $X$, i.e.
$\int_X x \cup y$. Again it is convenient to write it as
$\Omega(x)=(-1)^{h(x)}$ for an integer-valued function that is
defined mod 2.

\vspace{3mm}
Defining $\Omega(x)$ at the level of $K$-theory automatically includes
forms of all odd dimensions \cite{Wi4}. Analogously to the IIA   
construction, we need $x \otimes {\bar y}$, which is in
$K^2(X)={\widetilde K}^0(X \times S^1 \times S^1)$, and define
\(
(x,y)= \int_{X \times S^1 \times S^1}
{\hat A}(X\times S^1 \times S^1) ch(x \otimes {\bar y}).
\)

\vspace{3mm}
There is a subtlety in defining $\Omega(x)$ \cite{Wi4}. Replacing the
element
$x \otimes {\bar x}$ of ${\widetilde K}^0(X \times S^1 \times S^1)$ 
by its complex conjugate amounts to exchanging the two $S^1$ factors.
In addition, this element is trivial if restricted to  
$X\times S^1 \times \{*\}$ or $X\times \{*\}\times S^1$. The two
properties
amount to saying that $x \otimes {\bar x}$ can now be interpreted
as an element of the Real group $KR(X\times S^2)$, with the involution
given by a reflection on one of the coordinates of $S^2$.
\footnote{By collapsing $S^1 \times \{*\}$ and $\{*\}\times S^1$, one maps
$S^1 \times S^1$ to $S^2$ and the reflection on one coordinate in $S^2$
is inherited from the map that exchanges the two factors of $S^1 \times
S^1$.} Now, by Bott periodicity of $KR$,
\(
KR(X\times S^2) \cong KO(X)
\)
which means that the element $x \otimes {\bar x}$ maps
to an element $w \in KO(X)$ and then one can define \cite{Wi4}
$h(x)=j(w)$, where $j(w)$ is the mod 2 index of the Dirac operator
with values in $w$, which has no elementary formula in general, but
for bundles $w$ whose complexification is of the form $x \otimes {\bar 
x}$,
then
\(
j(w)= \int_X {\hat A}(X) ch(w) \quad {\rm mod} 2 .
\)

\vspace{3mm}
Having reviewed Witten's construction in $K$-theory, we next look at the
twisted case, for which there is an analogous story. 
Since $H_3$-twisted 
$K$-theory 
is not ultimately the theory we are after for solving the S-duality 
problem, we 
will not attempt the whole construction but limit ourselves to certain
aspects that follow rather directly from the construction in the case
of $K^0$ in type IIA \cite{HM}, which in turn is a generalization 
of \cite{DMW} and \cite{MS} to the (nontrivial) $H_3$-twisted case.

\vspace{3mm}
In the presence of $H$-flux, the Ramond-Ramond fields
${{F}}$ are determined by
the twisted $K$-theory classes $x \in K(X, H)$ via the twisted Chern map
\(\label{F1}
\frac{{F}(x)}{2 \pi} = ch_{H}(x) \sqrt{{\hat A}(X)} \in
H^{odd}(X, H).
\)

As in the $K^0$ case \cite{HM}, the conjugate of $x$, 
$\bar x \in K^1(X, -H)$
\(\label{F2}
\frac{{{F}(\bar x)}}{2 \pi} = ch_{-H}(\bar x) \sqrt{{\hat A}(X)}
\in H^{odd}(X, -H)
\)

The RR field equations of motion can be written as
\(
d{{F}}=H_3 \wedge {F}
\)
which, at the level of differential forms says that
the RR fields determine
elements in twisted cohomology,   $H^{odd}(X, H)$, and 
at the level of cohomology this implies $H_3 \wedge {{F}}_{n}=0$.

\vspace{3mm}
We again have a diagram (analogously to \cite{HM}, and which was also 
essentially used in \cite{BEM})
 
\begin{equation} \label{rootA}
\begin{CD}
K^1(X, H)  \times K^1(X, -H)  @>>> K^0(X)
@>{\rm index}>>\ZZ \\
          @V{ch_H}\times {ch_{-H}}VV          @VV{ch} V      @VV{||}V \\
H^{odd}(X, H)  \times H^{odd}(X, -H)    @> >>
H^{even} (X)   @>\int_X{\widehat{A}(X)}\wedge>>\ZZ
\end{CD}\end{equation}
where the upper row contains 
the cup product pairing in twisted $K$-theory
followed by the standard index pairing of elements of $K$-theory
with the Dirac operator (and we have used Bott periodicity).
The bottom
horizontal arrows are cup product in twisted cohomology
followed by cup
product by ${\widehat{A}(X)} $
and by integration. By the Atiyah-Singer index theorem, the diagram
\eqref{rootA}
commutes.  Therefore the normalization given to the Chern character in
the definition of
$\frac{{{F}(x)}}{2 \pi} $
makes the pairings in twisted $K$-theory and
twisted cohomology isometric.

\vspace{3mm}
Again here there is the subtlety in the self-duality
$*{{F}}={{F}}$, which would be resolved \cite{Wi4, MW} by interpreting 
this 
self-duality as
a
statement in the quantum theory and summing over half the fluxes, i.e. over
a maximal set of commuting periods. 
The lattice is $\Gamma_{K_H}=K^1(X,H)/K^1(X,H)_{tors}$. This is isomorphic
to the image of
the modified Chern character homomorphism of ${\Z}_2$-graded rings,

\(
\sqrt{\hat{A}(X)} \wedge ch_H : K^1(X,H) \rightarrow H^{odd}(X,H;\R)  
\label{esqrt} 
\)
(via a similar arguement as the untwisted case)
and the kernel is $K^1(X,H)_{tors}$, the torsion subgroup.
However, in this case, it would be desirable (to see whether 
it is possible) to  
construct the structures on the lattice 
in analogy to the twisted IIA case \cite{HM}.


\vspace{3mm}
One possible approach to trying to understand the twisted case
better is to have a better understanding of the twisted backgrounds.
This may include understanding the role of {\em higher twisting}.
Higher twistings have been used in the work of Freed, Hopkins and
Teleman on the Verlinde algebra \cite{turk} \cite{FHT}.
For complex $K$-theory, 
by \cite{Tel}, a twisting of complex $K$-theory over X is a principal 
$BU_{\otimes}$-bundle
over that space. By (\ref{BU}), this is a pair $(\delta, \tau)$   
consisting of a {\it determinant twisting} $\delta$, which is a
$K(\ZZ,2)$-bundle over X, and a {\it higher twisting} $\tau$, which is
a $BSU_{\otimes}$-torsor, so that twistings are classified, up to 
isomorphism,
by a pair of classes $[\delta] \in H^3(X, \ZZ)$ and $[\tau]$ in the
generalized cohomology group $H^1(X,BSU_{\otimes})$. This group is 
not very simple to deal with over $\ZZ$, but becomes more manageable
over $\QQ$, where it coefficient $BSU_{\otimes}$
becomes a topological abelian group, isomorphic to
$\prod_{n \geq 2}K(\QQ,2n)$ via the logarithm of the Chern character
$ch$. In this latter case, Teleman then asserts \cite{Tel} that the 
twistings of rational 
$K$-theory over X are classified, up to isomorphism, by the group 
$\prod_{n> 1}H^{2n+1}(X,\QQ)$.

\vspace{3mm}
One place where it seems that higher twisting of $K$-theory does show
up is in considering non-abelian spacetime.
It is a well known fact that the $B$-field, whose field strength is $H_3$,
measures non-commutativity of spacetime. In this section, we 
propose why higher twisting may be relevant from
this geometric point of view. 

\vspace{3mm}
First of all, mathematically, an infinitesimal deformation of any structure is measured
by its first Quillen cohomology \cite{q}. Now for example flat spacetime
is characterized by its ring of functions, which is (up to some completion)
roughly a polynomial algebra. Now in the category of rings, $1$st
Quillen cohomology is $2$nd Hochschild cohomology, which, for the
polynomial algebra $R$, is simply the space of possible violations of
commutativity, i.e. the space of antisymmetric matrices over $R$, i.e.
$B$ fields.

\vspace{3mm}
This predicts that perhaps the string moduli space should contain 
backgrounds whose spacetimes have spaces of functions which are global
non-commutative rings. We do not know if the IIA and IIB theory can indeed
be fully recovered for such backgrounds, although it is worth noting
that $K$ theory of non-commutative topological rings makes 
sense. Using Serre-Swan theorem one can 
define for a not necessarily commutative ring $A$, $K_0(A)$ as the group 
defined by the semigroup of isomorphism classes of finite projective 
A-modules. The definition of $K_1(A)$ follows the same pattern as the 
commutative case, provided $A$ is a Banach algebra. Bott periodicity 
extends to all Banach algebras.

\vspace{3mm}
There is also an intermediate step, i.e. ``perturbative'' non-commutative
geometry in the sense of Connes \cite{con}. This is defined in a similar
fashion as a super-manifold: a manifold $X$ together with a ``structure sheaf'',
or a bundle of non-commutative algebras. Atiyah and Segal \cite{AS}
comment that while a $B$-field is precisely the data needed
by twisting $K$-theory cohomologically, a $B$-field
realized by such non-commutative algebra bundle gives therefore
Connes' non-commutative geometry. To have a bundle of non-commutative
algebras, however, we first have a bundle of their indecomposibles,
i.e. a vector bundle, which defines an element of
$K$-theory.

\section{The Diaconescu-Moore-Witten anomaly}
\label{DMWa}

Here we show that there is no Diaconescu-Moore-Witten anomaly in the 
ordinary untwisted IIB
partition function.
In computing the quadratic structure $\Omega$, we took a class $x$ in
$K$-theory ($K^1$ in case of IIB), and by multiplying it by its conjugate,
created a class in real $K$-theory ($KR^{1+\alpha}X\cong KR^0X$ in IIB)
and then capped with the KR theory fundamental class to get the mod 2
index. Witten's method for identifying the anomaly is to choose a
polarization of the form $\omega$, i.e. a maximal isotropic space.
On such space $W$, there exists a characteristic, i.e. an element
$\theta$ such that for $x\in W$, $\Omega(x)=\langle x,\theta \rangle$   
(considering $\Omega(x)\in \Z/2$).

\vspace{3mm}
Now Witten takes a polarization compatible with the AHSS.
For IIA, this is the dual of $K$-theory elements
supported on the $k$-skeleton, for $k>5$, in the AHSS. 
Witten then argues that the characteristic
must be supported in the AHSS in dimension four. 
The filtration degree in real AHSS we need to get to is 8
(to apply the mod 2 index, which lands in $KO^2$), so
a priori one can take the characteristic to be supported
in filtration degrees $\geq 4$ (which is, from the point of view
of this pairing, dual to $W$).

\vspace{3mm}
The same subtle argument can be applied
to IIB. Of course, there is no canonical polarization, but
we can take simply some isotropic subspace of the $\Omega$-dual
of $K^1$-theory elements supported on the $>5$-skeleton in the AHSS.
This amounts to just selecting some polarization of the part
supported in cohomological dimension 5. Now for such choice of $W$,
although not canonical, this time the characteristic can
be taken supported in filtation degrees $\geq 5$. The point is,
again, the mod 2 index is supposed to end up in $KO^2$, but
we also apply the $1+\alpha$ shift. So, it seems that even before
the shift, we want the complementary degree in the (cohomology)
AHSS to be $-8$, so $-4$ for one class $x$, which occurs in filtration 
degree $5$.

\vspace{3mm}
Then the only non-trivial dimension is $5$. So the question will be
analogous to IIA: Can there be a class $a=\theta$ in $H^5(X,\ZZ)$ which
supports a differential in the AHSS? This is not possible, as we
argued before: if $d_3 a=b$ then choose $c\in H^2(X,\ZZ)$ with   
$b\cup c$ not divisible by 2 (Poincar\'e duality). But then $c$ must
support a $d_3$ in the AHSS but that cannot happen, because second
cohomology is represented by maps to $CP^\infty$, hence any second
cohomology class lifts to $K$-theory.
Therefore there is no Diaconescu-Moore-Witten anomaly in the ordinary
$10$-dimensional untwisted IIB partition function.

\vspace{3mm}
Note that this observation does not violate ordinary untwisted T-duality
in the canonical case: when $X_{10}=X_9\times S^1$ and everything
is $Spin$, then one must
have $W_7(X_{10})=W_7(X_9)=0$ for the usual reason that $\beta Sq^6$ when
applied to the fundamental homology class of $X_9$ would have to be
an integral homology class in dimension $2$, which is again impossible
by Poincar\'e duality.

\section{S-duality and $K$-theory}
\label{S}
\subsection{The puzzle}

Diaconescu, Moore
and Witten \cite{DMW} discuss the apparent incompatability of
twisted $K$-theory and S-duality. In order to have an $SL(2,\ZZ)$
symmetry, the monodromies of $\tau$ should be trivial, which implies that
$F_1$, which determines the monodromies of $Re\tau$ should vanish (recall
that $\tau=C_0 + ie^{-\phi}$). So the condition for anomaly cancellation
(i.e. the $W_3$ story \cite{FW}, see also \cite{KS}) for the next 
nontrivial field 
$F_3$ is  $(Sq^3 + H_3)\cup F_3=0$.
Now the pair $H^i$ is expected to transform in the two-dimensional  
representation of the modular group. First, $\tau \rightarrow \tau +1$, 
sending $F_3$ to $F_3+H_3$ and $H_3$ to itself, encounters no problems 
since 
$H_3 \cup H_3=Sq^3 H_3$
which implies that the anomaly equation is invariant. But it is not
invariant under the full group, e.g. sending $F$ to itself and $H$ to
$H_3+F_3$. The
$SL(2,\ZZ)$-invariant extension of the above equation is \cite{DMW}
\(
F_3 \cup H_3 + \beta Sq^2 (F_3 + H_3)=0.
\label{Sq}
\)
One immediate question is that of justification (and interpretation)
of the term
\beg{esqb}{\beta Sq^2H_3=H_3\cup H_3.}

\vspace{3mm} 
The problem can be restated as finding an S-duality extension
of $d_3$ in AHSS (see \cite{EV} for discussions on this 
point). There have 
been several proposals
concerning this.
The authors of \cite{EV} state that
they have verified equation (\ref{Sq}) for IIB configurations obtained   
by T-duality of M-theory on a torus, and proposed a nonlinear extension
of the differentials. In \cite{DFM} this problem was also addressed,
focusing on the torsion components of the
fields, starting from M-theory on a two-torus. Using their cubic
refinement
law for the M-theory phase and the quadratic refinement of the mod 2
index $f$, Diaconescu, Freed and Moore \cite{DFM} also derived the   
$SL(2,\ZZ)$-invariant equation of motion for the torsion fields,
\(
Sq^3 (F_3)+Sq^3 (H_3) + F_3 \cup H_3 = P,
\)
where $P$ depends only on the topology and the spin structure of the
nine-manifold, and is defined by
\(
e^{i\pi f(a)}=\langle a, P\rangle.
\)

\vspace{3mm} 
Let us now attempt an explanation of the \rref{esqb} term in the equation
\rref{Sq}. First of all, recall say from \cite{AS}, Proposition 4.1, how
one derives the differential $d_3$ in the twisted Atiyah-Hirzebruch
spectral sequence: the term \rref{esqb} must be explicitly excluded,
and the argument for excluding it is that one should have, in the
twisted $K$-theory AHSS,
\beg{esq0}{d_3(0)=0.}
Therefore, the presence of the term \rref{esqb} indicates that \rref{esq0}
must be violated in the $K$-theoretical intepretation of IIB. In other
words, the equation \rref{Sq}, for a given $H_3$, is not linear in $F_3$
and this is forced if we want to have modularity between the $H_3$ and
$F_3$ fields. 

\vspace{3mm}
In what way do we have to modify twisted $K$-theory to allow for a violation
of \rref{esq0}? First, let us recall which twisted generalized cohomology
theories we call twisted $K$-theory. A twisted generalized cohomology theory
is a bundle of cohomology theories on $X$. So, for twisted $K$-theory,
we would like the fiber over each point of $X$ to be $K$-theory. However,
this condition is not enough. We would get a lot more ``twisted $K$-theories''
if we only imposed that condition. The salient point is that 
we also insist that twisted $K$-theory be a {\em module} over $K$-theory,
which forces the ``structure group'' of the bundle of $K$-theories in
question to be the multiplicative infinite loop space $GL_1(K)$ of
$K$-theory. We have (see e.g. \cite{Tel}) 
\(
GL_1(K)=BU^{\otimes} \sim \ZZ/2\ZZ \times {\CC P}^{\infty} \times BSU
\label{Ktwist}
\)
Aside from the twistings coming from the group
$\{\pm1\}$, the splitting (\ref{Ktwist}) refines to a decompostion of
the spectrum $BU_{\otimes}$ of 1-dimensional units in the classifying
spectrum for complex $K$-theory \cite{MST}
so that one has the factorization 
\(
BU_{\otimes} \cong K(\ZZ;2) \times BSU_{\otimes}.
\label{BU}
\)

\vspace{3mm}
Now if we have a violation of \rref{esq0}, clearly we cannot
be dealing with a $K$-theory module. Can we then have
some further generalized twisting, where, for a particular $H_3$,
the choice of allowable $F_3$'s would not form a vector space? 
In other words, could one consider a form of twisted $K$-theory
which is not a module cohomology theory over ordinary $K$-theory?

\subsection{Generalized $K(\Z,3)$-twisted K-theory, and homotopy theory}

In this section, we will show that the kind of
theory described in the end of the last subsection,
also cannot exist if the twisting space
is $K(\Z,3)$. However, first we
have to understand what its existence would mean
in terms of homotopy theory. So far, we
have only exhibited one differential, or obstruction, \rref{Sq}
for the cohomological pair $(H_3, F_3)$ to lift to the theory.
For the theory to actually exist, we would have to exhibit a 
{\em classifying space}, i.e. a topological space $\mathcal{B}$
such that our affine-twisted $K^1$-group would be classified by homotopy
classes of
maps
\(
X\r \mathcal{B}. 
\)
As it turns out, such space cannot exist. To understand this, the reader should
review the Appendix for the construction of the space classifying $K^{1}_{tw}$.
\footnote{When not referring to a particular space, we use the notation
$K_{tw}$ for twisted K-theory.}
We see then that the space $\mathcal{B}$ should sit in a fibration
\footnote{Here by fibration, or more precisely a fibered sequence, we mean a sequence
of spaces
$$F\r E\r B$$
where $E\r B$ is a fibration, which means
roughly that $E$ is fibered without singularities over $B$
(more precisely that it satisfies the homotopy lifting property), and the fiber is $F$.
Any map in topology may, up to homotopy, be replaced by a fibration, so this
places no homotopical restriction on the map $E\r B$.}
\beg{egp1}{SU\r \mathcal{B}\r K(\Z,3).}
Then its Postnikov tower would have again the factors \rref{epostk1}, but
the lowest Postnikov invariant should now be given by the equation
\rref{Sq} instead of \rref{Sqk}.

\vspace{3mm}
Now applying the loop functor $\Omega$ ($\Omega X$ is the space of
all maps $S^1\r X$ preserving selected base points) to \rref{egp1}, 
and taking suitably rigid models, we get an extension of topological groups
\beg{egp2}{BU\r \Omega B\r \C P^{\infty}.
}
We will investigate $\Omega B$ as a homotopy associative $H$-space
\footnote{Recall that a $H$-space is a topological group up to homotopy.}.
First, since $BU$ is commutative up to homotopy, \rref{egp2} gives
rise to a well defined action up to homotopy
\beg{egp3}{\C P^{\infty}\times BU\r BU.}
Now actions up to homotopy of the form \rref{egp3} can be classified.
For our purposes, it suffices to consider actions which act through
$H$-space homomorphisms, i.e. preserve the group structure of $BU$
up to homotopy.

\vspace{3mm}
Now maps $BU\r BU$ are characteristic classes in $K$-theory, among
which the additive ones are linear combinations of the Segre classes
(otherwise known as Adams operations) $\psi_n$, which take a line bundle
to its $n$'th power. Note that the composition of Adams operations is
$$\psi_m\circ\psi_n=\psi_{m+n},$$
so if we write $a^n$ for $\psi_n$, one can write \rref{egp3} as
a power series with integral coefficients in two variables
$$\phi(x,a),$$
and the condition that \rref{egp3} is an action up to homotopy reads
\beg{egp4}{\phi(x,a)\cdot\phi(y,a)=\phi(x+y+xy,a).}
To make this clearer, note that after rewriting formally
$$\phi(x,a)=\Phi(1+x,a),$$
\rref{egp4} becomes
$$\Phi(1+x,a)\cdot\Phi(1+y,a)=\Phi((1+x)(1+y),a)$$
to which clearly the only solutions are
\beg{egp5}{\phi(x,a)=\Phi(1+x,a)=(1+x)^m,\; m\in\Z.}
Therefore, $\phi$ cannot depend on $a$. What we have proven is that
at least up to homotopy, the only actions \rref{egp3} are 
ordinary twistings of $K$-theory
by $m$'th power of the line bundle. Moreover, to get the exactly the multiplicative
term in the Postnikov invariant \rref{Sq}, we must have $m=1$.

\vspace{3mm}
We now continue our classification of the homotopy $H$-spaces \rref{egp2}.
Now that the action \rref{egp3} has been identified, extensions \rref{egp2}
up to homotopy are classified by a cohomology group which we can roughly write
as a ``group cohomology''
\beg{egp6}{H^2(\C P^{\infty},BU)}
where $\C P^{\infty}$ acts on $BU$ by the action \rref{egp3}, which is
\rref{egp5} with $m=1$. More precisely, if $[X,Y]$ denotes homotopy classes
of maps from $X$ to $Y$, then \rref{egp6} is a cohomology group
\(
[\C P^{\infty},BU]~ {\buildrel {d_1} \over \longrightarrow}~
[\C P^{\infty}\times\C P^{\infty},BU]~ {\buildrel {d_2} \over 
\longrightarrow}~
[\C P^{\infty}\times \C P^{\infty} \times \C P^{\infty},BU].
\)
Using the fact that $K^0((\C P^{\infty})^{\times n})$ is the ring of
integral power series in $n$ variables (copies of the canonical complex
orientation of $K$-theory),
we may write an element in the middle term as a power series with integral
coefficients
$$f(x,y),$$
and then using \rref{egp5} for $m=1$, we have
$$(d_2f)(x,y,z)=f(x+y+xy,z)-f(x,y+z+yz)+f(x,y)(1+z)$$
and similarly 
$$(d_1g) (x,y)=g(x+y+xy)-g(x)(1+y).$$
As before, things become clearer when we formally rewrite
$$f(x,y)=F(1+x,1+y),$$
$$g(x)=G(1+x).$$
The differential then becomes
$$d_2f=F((1+x)(1+y),(1+z))-F(1+x,(1+y)(1+z))+(F(1+x,1+y))(1+z),$$
$$d_1g=G((1+x)(1+y))-(G(1+x))(1+y).$$
This however is the standard pattern of the cobar construction with repeat
term, which has no higher cohomology: indeed, if 
$$d_2f=0,$$
then
\beg{egp7}{f=d_1g}
where
$$g(x)=G(1+x)=F(1,1+x)=f(0,x).$$
To see this, calculate
$$d_1g=F(1,(1+x)(1+y))-(F(1,1+x))(1+y),$$
while
$$0=d_2f(1,x,y)=F(1+x,1+y)-F(1,(1+x)(1+y))+(F(1,1+x))(1+y),$$
thus proving \rref{egp7}. 

\vspace{3mm}
We have therefore shown that the cohomology group \rref{egp6} vanishes,
so at $m=1$, ordinary twisted $K$-theory is the only candidate for $\Omega 
\mathcal{B}$ in the extension \rref{egp2} as homotopy associative 
$H$-spaces. Now recall that when we apply 
the loop functor $\Omega$ to {\em any} fibration of topological spaces
\rref{egp1}, we obtain (upon selecting an appropriately rigid model)
an extension of topological groups \rref{egp2}, and that this extension
moreover determines the fibration \rref{egp1}. Now we have classified
the extension \rref{egp2} on the level of $H$-spaces, which is
weaker than topological groups.
A priori, this does not exclude the possibility that
some subtle higher invariant could further distinguish between the
extensions of topological groups \rref{egp2}. However, the {\em first} Postnikov
invariant of $\mathcal{B}$ (the one which goes from dimension $3$ to
dimension $6$) is determined by the structure of $\Omega\mathcal{B}$
as homotopy associative $H$-space. Therefore, we have shown that 
\rref{Sq} cannot occur as first Postnikov invariant of a space sitting
in the fibration \rref{egp1}. In other words, no version of $K(\Z,3)$-twisted
$K$-theory, i.e. no theory which would be classified by a space 
$\mathcal{B}$ sitting in a fibration sequence of the form \rref{egp1}, can
be compatible with S-duality, which is needed to solve the puzzle of \cite{DMW}.

\subsection{Even more general twisted K-theory?}
We have shown that no generalized $K(\Z,3)$-twisted K-theory solves the 
problem of producing a theory
which would explain the equation \rref{Sq}. 
>From a physics point of view, one might ask whether 
any kind of higher-twisted K-theory might solve the problem. In particular, 
it seems reasonable to ask whether K-twisted K-theory should
play a role. If so, then that would imply a symmetry between 
the RR fields and the NSNS fields. Let us discuss how one might
form ``NSNS fields'' in odd degrees. Of course, we have the 
standard field $H_3$. Aside from $H_7$ the dual of $H_3$, 
one can also include the field 
strength $de^{\varphi}$ of the dilaton $\varphi$, which is of degree
one, its dual, and also the five-form field strength since it is 
duality-symmetric.
One then has forms of all odd degrees less than
ten, and one might then argue, in analogy to the situation with the 
RR fields, that they form 
a total Neveu-Schwarz field
\(
H_{NS}=\sum_n H_{2n+1}
\)
where $H_{2n+1}$ are the individual fields identified above. This 
might then be viewed as a K-theory element which provides the higher
twisting for the Ramond-Ramond fields, since as we saw earlier, 
rationally are of the form \cite{Tel}
\(
\prod_n H^{2n+1}(X,\QQ).
\)

But where would one look for such theory? First of all, recall 
again that 
ordinary higher-twisted module $K$-theory
is not really
$K$-twisted, but $gl_1(K)$-twisted,
which involves the factors $K(\ZZ,2) \times BSU$,
so using this, one still does not get the \rref{Sq} equation. 
Also, the self-dual $F_5$  
in fact places doubts on any hope that a $(K,K)$-approach to IIB 
could give S-duality: Bott periodicity, which is intrinsic to $K$-theory,
has a physical interpretation in tachyonic condensation where it shows
up as a Gysin map. This periodicity relates all the fields $F_{2i+1}$.
Now there is no obvious physical argument why the NS-NS-fields
should be related by such construction. Even if this were the
case, one would have to explain how come the construction passes
through the self-dual $F_5$ field and returns back to NS-NS fields. 

\vspace{3mm}
It is difficult to make a pursuasive argument that
{\em no} twisting of $K$-theory could 
possibly work. The reason is that it is difficult
to specify what exactly one would mean by the most general
kind of twisting. For example, Zhang \cite{Z}
defines $K$-twisted $K$-theory as cohomology of a complex
$$\diagram
\rto^\alpha &K^0(X)\rto^{\alpha}& K^1(X)\rto^{\alpha} &K^0(X)\rto^{\alpha}&
\enddiagram
$$
where $\alpha\in K^1(X)$ is a class, and makes calculations with
this construction. Now although a theory of this
type can lead to valuable calculations, even defining it 
however depends on calculational properties of $K^*(X)$
(one must have $\alpha^2=0$ -- a priori it is only $2$-torsion, which
is precisely where the subtleties lie with the equation \rref{Sq}).

\vspace{3mm}
In homotopy theory, the desirable property would be that our hypothetical theory,
let us call it $\mathcal{K}^1$
should be representable, which means one should have
$$\mathcal{K}^1(X)=[X,\mathcal{B}]$$
for some space $\mathcal{B}$. Now one may ask what type of space we allow
as $\mathcal{B}$. Further, if we are referring to a twisting of $K$-theory,
we should at least have a fibration
\beg{efibx}{SU\r\mathcal{B}\r T
}
for some space $T$. In the previous section, we have excluded $T=K(\Z,3)$
as a candidate. But what most general choice of $T$ should we allow?
A homotopy theorist might answer $T=BG$ where $G$ is the monoid
of self-maps $BU\r BU$ either as a space or as an infinite loop space.
However, one could go even more general. For example, one could argue
that $K$-theory of $K$-theory should also be considered a twisting of
$K$-theory. This was calculated by Baas-Dundas-Rognes \cite{bdr}, and
shown to be related to elliptic cohomology, which we will consider
in the next section.

\vspace{3mm}
In general, homotopy theory allows us to say this: since we have the
field $H_3$, and can neglect lower dimensions for the purposes of
deriving our condition, we
can assume the space $T$ of
\rref{efibx} is $2$-connected and has
\beg{efiby}{\Z\subset \pi_3(T).}
We should be more precise here: of course we are interested in situations
where spacetime $X$ would not be simply connected. So, the most general
kind of fibration we should consider is in fact
\beg{efibgen}{U\r \mathcal{B}\r T.}
However, for the purpose of merely deriving a condition on \rref{efibgen},
we may take a $2$-connected cover of all terms of \rref{efibgen}, and 
still have a fibration sequence. Then we get \rref{efibx}
with $T$ $2$-connected.

Now we can then ask what Postnikov invariants (also called
$k$-invariants) are attached to the
generator \rref{efiby}?
By the result of the last section, at least one of those 
$k$-invariants must be non-zero. Now assuming $\pi_4(T)=0$ (since we do
not see an obvious $4$-dimensional field),
the lowest possible $k$-invariant of $T$ is
$$k_3(H_3)\in H^6(K(\Z,3),\pi_5(T)).$$
But if this $k$-invariant is non-zero, it alters, by definition, the
equation \rref{Sq}, because the target of the $k$-invariant
considered there is $H^6(K(\Z,3),\pi_5(SU))$, which is a different
term of the fibration \rref{efibx}. Assuming, however, that the lowest
$k$-invariant of $H_3$ is in $H^n(K(\Z,3),\pi_{n-1}(T))$ for $n>3$,
we see that the space $T$ cannot be a form of $K$-theory or ordinary cohomology,
since in those theories, all classes with non-trivial $k$-invariants
have a non-trivial $k_3$. We see therefore that one way or another,
we are led to some type of other theory than $K$-theory or ordinary
cohomology.

\section{Elliptic cohomology}
\label{vs}

Let us now drop the restriction that the theory
we are searching for be a priori a twisting of
$K$-theory. Let us recapitulate
the requirements we have on a theory which used to calculate the IIB partition
function:

\vspace{3mm}
\begin{enumerate}
\item \label{ii1}
$S$-duality, i.e. modularity with respect to $H$, $F$

\vspace{3mm}
\item \label{ii2}
Higher generalized cohomology - at least a $2$'nd generalized cohomology
group must be defined in our theory to imitate the untwisted $K^1$-partition
function construction reviewed above

\vspace{3mm}
\item \label{ii3}
$T$-duality with IIA, which should in some form extend to the $>10$-dimensional
limits.

\end{enumerate}

\vspace{3mm}

In the last section, we concluded that \ref{ii1} cannot be satisfied by
any generalization of $K(\Z,3)$-twisted $K$-theory. However, 
\ref{ii2} casts doubt on the viability of any twisted generalized cohomology
approach: when the twisting space is introduced into the theory, an intrinsic
non-commutativity is introduced which seems to prevent further ``delooping''
of the theory into a modular second cohomology group. It therefore seems
that a candidate for the theory we are searching for, should be, after
all, an {\em untwisted} generalized cohomology theory.
But how could the twisting of $K$-theory untwist in another cohomology theory
which carries at least the same amount of information? There is
some evidence for 
an answer. Douglas \cite{dtwist} notes that twistings of $K$-theory
result in maps
$$K(\Z,3)\r TMF,$$
in fact defining ``elliptic line bundles''. Therefore, what looks like twisting
to the eyes of $K$-theory, untwists and becomes merely a multiplication
by a suitable element in $TMF$ or any suitable form of elliptic cohomology.

\vspace{3mm}
One can in fact push this a bit further: the geometric interpretation
of ordinary cohomology twisting of $K$-theory is as a gerbe 
\cite{BCMMS} \cite{BM}. However,
a gerbe should only be a special case of a $2$-vector space bundle.
The problem is that this intuition has not yet been made precise
mathematically: a $2$-vector space should be a lax module (in the
sense of \cite{hk1}) over the symmetric bimonoidal category $\C_2$ of
finite-dimensional complex vector spaces. Naively, a $2$-vector
space over $X$ would just be a bundle, assigning to each point $x$
of $X$, continuously a $2$-vector space $\mathcal{V}_x$. One could
then define $\mathcal{V}$-twisted $K$-theory simply as a the $K$-theory
of the category of continuous sections of $Obj(\mathcal{V}_x)$, as
$x$ varies. The problem with making this mathematically precise
is that the $2$-category of $2$-vector spaces is unexpectedly rigid,
and there are not enough equivalences to get good examples of $2$-bundles.
Solutions to this problem have been proposed (\cite{bdr,hk1}), but
it is fair to say that so far these structures have not been investigated
in detail. The main point is that it seems that a suitable ``group completion''
of $\C_2$ must be found, i.e. a category of super-vector spaces with well
behaved inverse. One mathematical way one can talk rigorously
about the $2$-vector space twistings of $K$-theory is via
the spaces $GL_n(K)$,whose limit determines the algebraic $K$-theory
of the $E_{\infty}$ ring spectrum $K$. In any case, from the work
of Baas-Dundas-Rognes \cite{bdr}, we expect that $2$-vector spaces, 
with perhaps some extra structure, also determine elliptic cohomology
(or $TMF$) elements, so this type of twisting by a higher-dimensional
$2$-vector bundle would also untwist in elliptic cohomology.

\vspace{3mm}
Another hint that twisted $K$-theory should lead
up to elliptic cohomology is the result of
Freed, Hopkins and Teleman \cite{FHT} relating the Verlinde algebra
of a compact Lie group
to its (ordinary cohomology-twisted) equivariant $K$-theory. Perhaps
by considering the full force of twisting, one could 
cohomologically explain the full
anomaly of the chiral WZW models? But anomaly of chiral CFT
is kind of ``$1$-st CFT cohomology'' information, and we think that
elliptic cohomology and ``CFT cohomology'' are related ({\it cf.} 
\cite{hk}).

\vspace{3mm}
In any case, we have now gathered some evidence that the
theory we are searching for, which would satisfy \ref{ii1}, \ref{ii2},
should be elliptic cohomology. In elliptic cohomology, the both fields
$H_3$ and $F_3$ play symmetrical roles, and twisting is replaced by
multiplication. Thus, the equation \rref{Sq} is replaced by 
requiring that both $H_3$, $F_3$ be represented by elements of
elliptic cohomology.
But this also dovetails with point \ref{ii3}:
In \cite{KS}, we proposed that a correction of the IIA partition function
in the case of the $M$-theory limit should be defined by elliptic cohomology.
It is therefore natural that the theory containing
the partition function of the $T$-dual IIB theory should also be 
elliptic cohomology.
There is in fact a physical argument why a {\em modular} theory should
involve the higher-dimensional high energy limits: $K$ theory predicts
periodicity of the RR-fields, but the dual NS-NS fields do not seem to
share such periodicity. Therefore, to have a modular picture, we need
to consider a theory which breaks the periodicity of RR-fields. However,
it seems that this periodicity is only broken when a strong-type coupling
limit leading to dimensional expansion of spacetime is introduced.

\vspace{3mm}
Also the fact that S-duality is a part of U-duality gives an argument
why we should be led to higher dimension in considering IIB modularity:
we have a path in the string moduli space connecting a theory to
its U-dual. In the case of IIA theory, it has been shown that
in the limit, the theory expands to $11$-dimensions, and therefore
the theory must be at least $11$-dimensional also on the path
in the string moduli space. In fact, dimensional expansion is
a kind of condition which should be generic in the moduli space.
Therefore, in the path in the moduli space connecting IIB with
its U-dual, we should also see $>10$-dimensional theories in the
interior of the path, even though both limit theories are $10$-dimensional
(we think, in fact, that the correct number of dimensions is
$12$; in future work, we shall obtain evidence confirming that proposal).

\vspace{3mm}
Let us recapitulate what we concluded up to this point: If we want to
understand S-duality in type IIB string theory, we must find a way
of twisting $F_3$ by $H_3$ which is compatible with S-duality.
To do that, we must look beyond any kind of $K(\Z,3)$-twisted $K$-theory.
One possible approach is to look for a generalized
cohomology theory where the twisting untwists, i.e. $H_3$ is just represented
by a class in the theory. Such theory is elliptic cohomology. However,
if we select elliptic cohomology as an approach to the IIB S-duality
puzzle, we have also expanded the theory. In other words, we are
no longer in the situation of classical IIB string theory, but in
some dimensional expansion of that theory. 

\vspace{3mm} 
There remains a question which exact type of elliptic cohomology
theory one should use. 
The trouble is that mathematics has not yet perfected the connection
between elliptic cohomology and geometrical or physical objects such
as gerbes or $2$-vector spaces. Many variations of constructions
originating from such objects can be made, leading to slightly different
elliptic cohomology theories. We do not have a definitive answer as to
which theory to use (a similar difficulty was encountered in \cite{KS}).
In algebraic topology, this difficulty is circumvented by observing
certain common features of elliptic cohomology theories, and considering
them therefore, in a way, all at once.

\vspace{3mm} 
The ``ultimate'' elliptic cohomology theory (which however is
no longer an elliptic cohomology theory in the standard definition) 
is $TMF$, 
and will pursue this point in a follow-up paper \cite{follow}.
In this paper, we shall simply show that 
complex-oriented elliptic cohomology theory we used in \cite{KS} can
be used as a kind of toy model. It
is easier to analyze mathematically, and in fact one can
build, at least at free field approximation, a physical
quantum theory based on elliptic cohomology
whose partition function is directly
analogous to the
$K$-theory partition function introduced in \cite{Wi4}, \cite{DMW}
which serves as an approximation of the type II partition function,
and which was reviewed above. The question is what exact role this
theory plays. We have concluded above that this theory serves as
partition function of some dimensional expansion of type IIB string
theory which has S-duality.

\vspace{3mm}
In \cite{Wi4}, \cite{DMW},
to correctly understand the phases necessary to define the type II partition
function, it was necessary to involve real $K$-theory in addition to
complex $K$-theory. Similarly, in the elliptic theory, to get correct
phases for its partition function, one must involve real elliptic cohomology.
Passage to a real form of a complex-oriented theory however entails
an obstruction. In the case of $K$-theory, it was shown in \cite{DMW}
that the correct obstruction in the untwisted case
is $W_7$. In Section \ref{DMWa} above, we have shown that in IIB, in the
untwisted case, there is no such obstruction.
We observed however in \cite{KS} that 
in real elliptic cohomology, the obstruction is stronger:
the $10$-dimensional compact spacetime $Spin$-manifold $X$ must 
satisfy the condition
\beg{eobst}{w_4(X)=0,}
which is a necessary and sufficient condition for orientation with
respect to real elliptic cohomology. A physical interpretation of this
seemingly foreign condition will be proposed below.

\vspace{3mm}
Now the partition function is getting a
$\theta$-function. Its construction in the IIB case is directly analogous
to \cite{Wi4}. Instead of $K^1(X)$, one starts with $E^1(X)$ where $E$
is complex-oriented elliptic cohomology. The construction proceeds
precisely analogously as in the $K^1(X)$ case. The only delicate
point which requires explicit discussion is the phase. There, too,
we have an analogy, but we must make sure all steps of the analogy
make sense.
We have the pairing in $E^1(X)$:
\beg{eep}{E^1(X)\otimes E^1(X)\r E^2(X)\r E^{-8}=E^0}
where the second map is capping with the fundamental orientation class
in $E_{10}(X)$. To construct a $\theta$-function, however, we
need a quadratic structure, for which, just as in the case of IIA, we
need to consider real elliptic cohomology: A product of an 
$x\in E^1(X)$ with itself can be given a real structure, which just
as in the case of $KR$-theory reviewed above, gives rise to an element
of
\beg{eep1}{\omega(x)\in E\R^{1+\alpha}X.}
But now capping \rref{eep1} with the fundamental orientation class in
$E\R_{10}(X)$, we obtain an element in $E^{\alpha-9}$. As remarked in
\cite{KS}, this is a $\Z/2$-vector space generated by the classes
$$v_{1}^{3n-1}v_2^{2-n}\sigma^{-4}a^2,\; n\geq 1.$$
Therefore, we get a quadratic structure depending on one free parameter,
which leads to a precise IIB analog of the $\theta$-function constructed
for IIA using real elliptic cohomology in \cite{KS}.

\section{The $w_4$ obstruction, IIA-strings and T-duality}
\label{T}

Why would there be a $4$-dimensional obstruction 
such as \rref{eobst} to 
construction of partition function in any type II string theory?
Such condition (on $\lambda$, the four-dimensional 
characteristic class of the tangent bundle equaling that
of the gauge bundle) is present in type I and heterotic string
theory, but not in $10$-dimensional type II theory. The direct answer is,
as we noted above, that we have left the confines of classical type II string
theory, and passed to its dimensional expansion. So what we see is that
we have an additional obstruction which must vanish in order for this
expansion to be exist. This is perfectly consistent. But why is
the obstruction exactly $w_4$?

\vspace{3mm}
In \cite{KS}, we saw that analysing the conjectured $M$-theory
limit of IIA string theory, the $W_7$ anomaly of Diaconescu-Moore-Witten
\cite{DMW} gets refined to $w_4$ (the obstruction
to orientability with respect to real elliptic cohomology)
or even $\lambda \in \ZZ/{24}$ (the obstruction to orientability with
respect to tmf). 
Now we observed above that the $W_7$ anomaly does not show up in ordinary
untwisted IIB-theory, but in the last section reached the conclusion
that elliptic cohomology (or $TMF$) should really be the theory 
describing the behavior of IIB, in which case the obstruction
$w_4$ (or $\lambda\in\Z/24$) reappears. How does this relate to T-duality?
What is then the role of a $4$-dimensional obstruction in IIB when 
there are no $4$-dimensional RR-fields in IIB? The answer, perhaps,
should be that the reason we see the same obstruction for all
string theories is that we are considering dimensional expansion,
which, as observed above, should be a generic phenomenon in the
string moduli space. In this expansion, perhaps, all string theories
should be unified (the moduli space should be connected), and
this is why the obstructions merge. What is, however, the right
number of dimensions?

\vspace{3mm}
Physically, one can argue that the two theories,
IIA and IIB, are not completely symmetrical in ten dimensions.
Note that certain topological terms on the M-theory (and consequently
the IIA-) side that do not come from corresponding terms in the
IIB side. For example, the one-loop topological term
\(
\int_{Y^{11}}C_3 \wedge I_8
\label{1loop}
\)
in M-theory, which reduces to the term
\(
\int_{X^{10}} B_2 \wedge I_8
\)
in type IIA, does not match a corresponding term in IIB
\cite{Das}. Instead, there are terms, that depend inversely on
the radius of the extra circle in nine dimensions, that make the matching
possible, and that tend to zero as the tenth dimension (up to IIB) is
decompactified \cite{Das}. So if topological terms are not traced back to 
similar
terms, then there could be a mismatch if one wants to look at the full
story in the context of the $E_8$ bundle in eleven (and twelve)
dimensions. This is because the term (\ref{1loop}) was an essential
ingredient in Witten's formulation of the phase of M-theory as
an $E_8$ index
\footnote{Of course, there is also the Rarita-Schwinger part.}.
A complete discussion of the effect of T-duality on the partition
function will perhaps involve the fermionic path integral as 
well as the one-loop corrections, as was found in \cite{MS} for
the case of IIA on $X_8 \times T^2$. This would be beyond the scope 
of this paper.

\vspace{3mm}
On the other hand, IIB should
have a $12$-dimensional $F$-theory limit
\footnote{In this section, the word ``limit'' is not meant in
the dynamical sense, but more in the geometric or topological sense.}. 
\footnote{We note that at present, referring to $F$-theory is a little
dangerous, since its precise physical form is not clarified.
In particular, at physical signatures, no Lorentz-invariant formulation
of $F$-theory is known so far. We shall return to this point
in future work. For now, however, let us neglect this difficulty,
and assume that a physically consistent $F$-theory has been developed.}
Then there should also be
$T$-duality which would somehow unify the $F$-theory limit
of IIB with the conjectured $12$-dimensional $F$-theory limit
of $M$-theory (hence also IIA), although we must note that necessarily,
both of the IIA and IIB limits should end up in quite different
sectors of $F$-theory:
For example, assuming there is an F-theory limit to IIB, the elliptic curve
which is the fiber must have completely Ramond spin structure, i.e.
no non-separating curve can bound. The point is, that is the only spin
structure on an elliptic curve which does not break modularity.   
So, the elliptic fiber must have Ramond spin structure in order to
preserve S-duality.

\vspace{3mm}
As noted, IIA should also have $12$-dimensional limits. On one hand, implicitly
or explicitly, one limit shows up in \cite{DMW} in the form of the 
manifold $Z$
whose boundary is the $11$-dimensional spacetime of $M$-theory,
which in turn is conjectured to be a strong coupling form of IIA.
On the other hand, if the IIA-theory has an F-theory limit whose fiber
is an elliptic curve, that fiber must have a spin structure which
contains NS non-separating curves, because in the $11$-dimensional
M-theory whose spacetime is an $S^1$-bundle over $X$, the fiber must have 
NS-spin structure. 

\vspace{3mm}
We see therefore that T-duality in the F-theory
limits of IIA and IIB cannot be straightforward in the sense that
it would simply keep the fiber intact. This is a possible explanation
why in the ordinary untwisted $10$-dimensional partition functions
of IIA and IIB, the $W_7$-anomaly shows up in IIA but not in IIB,
and also of the other apparent asymmetries between $10$-dimensional
IIA and IIB theories. Note however that T-duality in F-theory
may be the key to explaining the role of the $w_4$-obstruction
to the elliptic IIB partition function: Although we know little
about the nature of F-theory at this point, it should however contain
both the IIA and IIB fields. Therefore, it is no surprise that 
a $4$-dimensional
topological obstruction is relevant in any sector of F-theory.

\vspace{3mm}
There is also another, purely mathematical, intuition why
orientability with respect to $tmf$ should be relevant to string
backgrounds. Stolz and Teichner \cite{stolz} conjecture that $tmf$
could be constructed as moduli space of supersymmetric conformal
field theories. Such theories, in dimension $10$ (more precisely
$(9,1)$), determine
$10$-dimensional string backgrounds. Therefore, there should
be a canonical $tmf$-class on the moduli space of $10$-dimensional
backgrounds which should be the Witten genus. This means that we should
consider the Witten genus on spacetime manifolds, but we know that
the Witten genus is well behaved precisely for manifolds which
have a $tmf$-orientation.

\vspace{3mm}
The above issues deserve further investigation and we hope to revisit them
in the near future \cite{follow}.

\vspace{1cm}
{\bf Acknowledgements}\\
We thank Edward Witten for emphasizing the importance of the problem
and for discussion. We also acknowledge useful conversations with Jarah 
Evslin and Varghese Mathai.
H. S. would like to thank the Michigan Center for Theoretical Physics
for kind hospitality during the intermediate stages of this project.

\vspace{15mm}

{\Large \bf Appendix: The homotopical construction of $K^{1}_{tw}$}

\vspace{3mm}
We give here the construction of the classifying space $\mathcal{K}$
of ``ordinary'' twisted $K^1$-theory.
Recall \cite{macl} that a monoidal
category 
\footnote{A monoid is a group-like object that fails to be 
a group by lacking an inverse. It is also a semigroup (i.e. a set 
equipped with an associative product) with an 
indentity element.}
is a category with a bifunctorial operation $+$
which is associative and unital (not necessarily
commutative) up to natural isomorphisms
which must satisfy appropriate coherence diagrams. 
It is in fact appropriate to call a monoidal category a
{\em lax monoid}: the general rule of thumb is that for any algebraic
structure, a {\em lax} verion of that structure is a category on which
all the operations are defined and satisfy all their required identities
(e.g. associativity, unit, etc.) up to natural isomorphisms, which in
turn sit in commutative {\em coherence diagrams}. There is no point in
going to too much detail about that here, but roughly speaking, imagine
we can convert in the (strict) algebraic structure one expression into
another using the defining identities in two different ways. Then, in 
the lax structure, these two ways define a required commutative coherence 
diagram.

\vspace{3mm}
Now for
a monoidal category $\mathcal{C}$, its classifying space
$B\mathcal{C}$ then has the natural structure of a topological
monoid, of which we can then again take
a classifying space $\mathbb{B}B\mathcal{C}$. (We deliberately
use here different symbols $B$, $\mathbb{B}$
for the category and monoid classifying spaces, since
these are different construction, and $\mathbb{B}$
is in fact with respect to the monoidal structure obtained from
topological realization of the categorical operation $+$,
whereas $B$ is just done with respect to the structure
of categorical composition. 

\vspace{3mm}
Then the space $\mathcal{K}$ can be constructed as
\(
\mathcal{K}=\mathbb{B}B\mathcal{C}
\)
where the monoidal category $\mathcal{C}$ has as objects pairs
$(L,V)$ where $L$ is a $1$-dimensional complex vector space
and $V$ is a finite-dimensional complex vector space,
and as morphisms pairs of isomorphisms, and the operation is
\beg{eoptw}{(L,V)+(M,W)=(LM,LW+V).
}
The unit is $(\C,0)$ and to see associativity, we have
\(
(L,V)+(M,W)+(N,Z)=(LMN,LMZ+LW+V).
\)
In fact, there is further discussion one can have: The space $\mathcal{K}$
sits in a fibration
$$U\r \mathcal{K}\r K(\Z,3).$$
Taking fixed loops, we get a fibration
\beg{efibk}{
BU\times \Z\r \Omega\mathcal{K}\r \C P^{\infty},
}
but it is a well known fact that we can choose models of the spaces involved
such that \rref{efibk} is actually an extension of topological groups.
In any case, since $\pi_1(\C P^{\infty})=0$, the fibration \rref{efibk}
does not twist the connected components of the fiber $BU\times\Z$, so,
we may get a group extension by restricting the fiber to $BU\times\{0\}$:
\beg{efibk1}{
BU\r \Omega\mathcal{K}_0\r \C P^{\infty},
}
and taking $\mathbb{B}$, we obtain a fibration
\beg{efibk2}{
SU\r\mathcal{K_0}\r K(\Z,3).}
The Postnikov tower of this space $\mathcal{K}$ has stages
\beg{epostk1}{K(\Z,3)\times K(\Z,3), K(\Z,5), K(\Z,7),...}
and the first Postnikov invariant is the map
\beg{epostk}{K(\Z,3)\times K(\Z,3)\r K(\Z,6)}
which is given by the equation
\beg{Sqk}{\beta Sq^2 F + H\cup F}
where $H,F$ are the characteristic $3$-dimensional cohomology classes of
the factors of the left hand side of \rref{epostk}.


\end{document}